\documentclass[english,aps,prb,preprint]{revtex4}
\usepackage[T1]{fontenc}
\usepackage[latin9]{inputenc}
\usepackage{float}
\usepackage{amsmath}
\usepackage{graphicx}
\usepackage{esint}

\makeatletter

\providecommand{\tabularnewline}{\\}

\@ifundefined{textcolor}{}
{%
 \definecolor{BLACK}{gray}{0}
 \definecolor{WHITE}{gray}{1}
 \definecolor{RED}{rgb}{1,0,0}
 \definecolor{GREEN}{rgb}{0,1,0}
 \definecolor{BLUE}{rgb}{0,0,1}
 \definecolor{CYAN}{cmyk}{1,0,0,0}
 \definecolor{MAGENTA}{cmyk}{0,1,0,0}
 \definecolor{YELLOW}{cmyk}{0,0,1,0}
 }

\makeatother

\usepackage{babel}
\begin{document}

\author{Glade Sietsema}
\email{glade-sietsema@uiowa.edu} 
\author{Michael E. Flatt\'e}
\email{michael_flatte@mailaps.org}
\affiliation{Department of Physics and Astronomy and Optical Science and Technology Center, University of Iowa, Iowa City, Iowa
52242, USA}
\date{\today}
\title{Magnonic band structure of a two-dimensional magnetic superlattice}
\begin{abstract}
The frequencies and linewidths of spin waves in a two-dimensional periodic superlattice of magnetic materials are found, using the Landau-Lifshitz-Gilbert equations. The  form of the exchange field from a surface-torque-free boundary between magnetic materials is derived, and magnetic-material combinations are identified which produce gaps in the magnonic spectrum across the entire superlattice Brillouin zone for hexagonal and square-symmetry superlattices.
\end{abstract}
\maketitle

\section{Introduction}
Advances in the control of spin-wave propagation and dynamics\cite{Serga2010} have led to the demonstration of magnonic bose condensation\cite{Demokritov2006} and coupling of electronic spin currents to spin waves in hybrid systems\cite{Ando2011}. Such effects, along with theoretical proposals to electrically-control spin-wave properties\cite{Liu2011}, and theoretical suggestions of high-temperature operation with small switching energies, may provide the foundation for an information-processing technology based on spin waves\cite{Kostylev2005,Khitun2010}. Any such technology would benefit from magnetic materials with designed spin-wave dispersion relations, group velocities, and linewidths. A common method of designing such features is the fabrication of a superlattice of different constituent materials, used to design electronic band structures in semiconductor superlattices, and photonic band structures in dielectric superlattices. 

Here we focus on the effect of a two-dimensional superlattice of magnetic materials on the magnonic frequencies and linewidths, obtained from a reciprocal-space solution to the Landau-Lifshitz-Gilbert (LLG) equation\cite{Gilbert2004}. Infinite cylinders of one magnetic material are  embedded in a second magnetic material in a periodic arrangement corresponding to a two-dimensional square lattice or hexagonal lattice. Large gaps within the spin-wave spectrum are obtained when the exchange constants and saturation magnetization of the two materials differ greatly; thus the gaps are considerably larger for cylinders of iron embedded within yttrium iron garnet (YIG) than within nickel. For iron embedded in YIG we demonstrate the existence of a gap throughout the superlattice Brillouin zone in the magnon spectrum for both square and hexagonal-symmetry magnonic crystals. In photonic crystals such a feature forms an essential element of photonic band gap materials\cite{Yablonovitch1987,John1987}, and permits the control of spontaneous emission of emitters embedded within the photonic crystal; here similarly the spontaneous emission of magnons from a source such as a spin-torque nano-oscillator could be suppressed by embedding this spin-wave emitter in a fully-gapped magnonic crystal.

Central to the accurate calculation of dispersion curves associated with a superlattice is the proper treatment of the boundaries between the two magnetic materials. For a magnonic superlattice the exchange field that enters into the LLG equations is discontinuous at the boundary, and that discontinuity strongly influences the spin wave dynamics. Two distinct forms for this exchange field have been described in the literature\cite{Vasseur1996,Krawczyk2008,Tiwari2010}, although to our knowledge it has not been pointed out that these two forms provide dramatically-different solutions to the LLG equation. In Section~\ref{formalism} we present  an explicit derivation of the correct form of the exchange field, followed by spin wave frequencies and linewidths  for various magnetic material combinations in Section~\ref{results}. In Section~\ref{comparison} we show that solutions to the LLG equations for the incorrect form of the exchange field differ greatly from those for the correct form, and furthermore the incorrect solutions are incompatible with the spatial symmetry of the lattice.


\section{LLG Formalism for a Quasi-Two-Dimensional Magnonic Crystal\label{formalism}}

\begin{figure}[h]
\begin{center}

\includegraphics[width=\columnwidth]{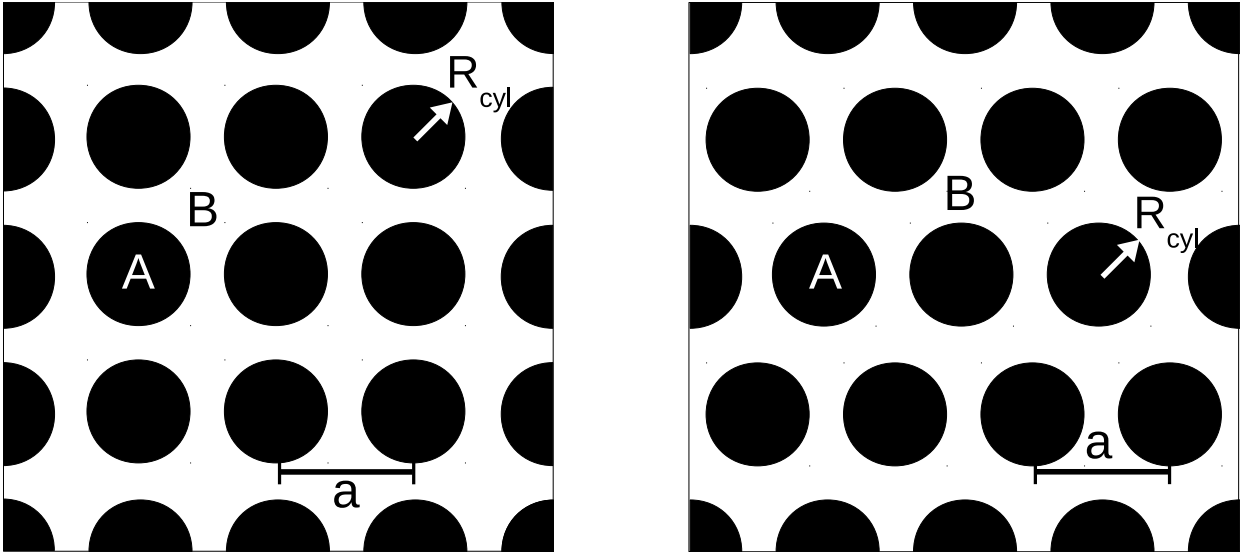}

\end{center}

\caption{Physical structure of the magnonic crystals studied here. The ferromagnetic
material B is the host for infinitely long cylinders of a different
ferromagnetic material A arranged in either a square (left) or hexagonal
(right) lattice. The lattice constant of the superlattice is $a$ and the cylinder radius is $R_{\rm cyl}$.}
\label{fig: Structure}
\end{figure}

We consider a magnonic crystal composed of an array of infinitely
long cylinders of ferromagnetic material A embedded in a second ferromagnetic
material B in a square or hexagonal lattice; the structures are shown in Fig.~\ref{fig: Structure}, and have lattice constant $a$ and cylinder radius $R_{cyl}$. The cylinders are aligned
parallel to 
a static external magnetic field $\mathbf{H_{\mathbf{\mathrm{0}}}} = H_0\hat z$, and the magnetization
of both materials is assumed to be parallel to $\mathbf{H_{\mathbf{\mathrm{0}}}}$.
The equation of motion for this system is the Landau-Lifshitz-Gilbert
(LLG) equation\cite{Gilbert2004}:
\begin{equation}
\frac{\partial}{\partial t}\mathbf{M}\left(\mathbf{r},t\right)=\gamma\mu_{0}\mathbf{M}\left(\mathbf{r},t\right)\times\mathbf{H}_{eff}\left(\mathbf{r},t\right)+\frac{\alpha\left(\mathbf{r}\right)}{M_{s}\left(\mathbf{r}\right)}\mathbf{M}\left(\mathbf{r},t\right)\times\frac{\partial}{\partial t}\mathbf{M}\left(\mathbf{r},t\right).\label{eq:LLG}
\end{equation}
Here $\gamma$ is the gyromagnetic ratio, $M_{s}\left(\mathbf{r}\right)$
is the spontaneous magnetization, $\alpha\left(\mathbf{r}\right)$
is the Gilbert damping parameter, and $\mathbf{r}$ is the three dimensional
position vector. The effective magnetic field 
\begin{equation}
\mathbf{H}_{eff}\left(\mathbf{r},t\right) = \mathbf{H}_{0} + \mathbf{h}\left(\mathbf{r},t\right) + \mathbf{H}_{ex}\left(\mathbf{r},t\right)
\end{equation}
acting on the magnetization $\mathbf{M}\left(\mathbf{r},t\right)$
consists of three terms: the external field $\mathbf{H}_{0}$, the
dynamic dipolar field $\mathbf{h}\left(\mathbf{r},t\right)$, and
the exchange field $\mathbf{H}_{ex}\left(\mathbf{r},t\right)$.

\subsection{Derivation of the Effective Electric Field}

We wish to derive the correct form of $H_{eff}\left(\mathbf{r},t\right)$
to use in Eq. (\ref{eq:LLG}) for our magnonic crystal. As shown by
Gilbert\cite{Gilbert2004}, the exchange field can
be obtained by taking the functional derivative of the exchange energy.
For a homogeneous material, the exchange energy is \cite{Kittel1949}
\begin{equation}
U_{ex}\left[\mathbf{M}\left(\mathbf{r}\right)\right]=\frac{A}{M_{s}^{2}}\int\left[\left(\nabla m_{x}\left(\mathbf{r}\right)\right)^{2}+\left(\nabla m_{y}\left(\mathbf{r}\right)\right)^{2}+\left(\nabla m_{z}\left(\mathbf{r}\right)\right)^{2}\right]d\mathbf{r},
\end{equation}
where A is the exchange stiffness constant. This yields the following
exchange field:
\begin{equation}
\mathbf{H}_{ex}\left(\mathbf{r}\right)=-\frac{1}{\mu_{0}}\frac{\delta U_{ex}\left[\mathbf{M}\left(\mathbf{r}\right)\right]}{\delta\mathbf{M}\left(\mathbf{r}\right)}=\frac{2A}{\mu_{0}M_{s}^{2}}\nabla^{2}\mathbf{M}\left(\mathbf{r}\right).
\end{equation}
For the inhomogeneous crystal considered here, the values of the exchange
constant and the spontaneous magnetization will differ for the
two ferromagnets, so $A$ and $M_s$ become spatially dependent quantities:
\begin{equation}
\begin{array}{c}
A\left(\mathbf{r}\right)=A_{B}+\Theta\left(\mathbf{r}\right)\left(A_{A}-A_{B}\right),\\
M_{s}\left(\mathbf{r}\right)=M_{s_{B}}+\Theta\left(\mathbf{r}\right)\left(M_{s_{A}}-M_{s_{B}}\right),
\end{array}
\end{equation}
where $\Theta\left(\mathbf{r}\right)=1$ in material A and $\Theta\left(\mathbf{r}\right)=0$
in material B. The  exchange energy for this inhomogeneous situation
is
\begin{equation}
U_{ex}\left[\mathbf{M}\left(\mathbf{r}\right)\right]=\int A\left(\mathbf{r}\right)\left\{ \left[\nabla\left(\frac{m_{x}\left(\mathbf{r}\right)}{M_{s}\left(\mathbf{r}\right)}\right)\right]^{2}+\left[\nabla\left(\frac{m_{y}\left(\mathbf{r}\right)}{M_{s}\left(\mathbf{r}\right)}\right)\right]^{2}+\left[\nabla\left(\frac{m_{z}\left(\mathbf{r}\right)}{M_{s}\left(\mathbf{r}\right)}\right)\right]^{2}\right\} d\mathbf{r},\label{eq:Inhom_Exch_Energy}
\end{equation}
By approximating the energy with $U_{ex}$ we have neglected non-exchange terms that would give rise to a surface torque (such as terms in the energy associated with surface-induced magnetic anisotropy). 

The total magnetization will consist of both a time-dependent term
and a time-independent term: $\mathbf{M}\left(\mathbf{r},t\right)=M_{s}\left(\mathbf{r}\right)\hat{z}+\mathbf{m}\left(\mathbf{r},t\right).$
Using the linear magnon approximation we assume that the time-dependent
magnetization  is small compared to $M_{s}\left(\mathbf{r}\right)$
and therefore we only keep terms up to first order in $\mathbf{m}\left(\mathbf{r},t\right).$
With these assumptions, the inhomogeneous exchange field derived from
Eq. (\ref{eq:Inhom_Exch_Energy}) is
\begin{multline}
\mathbf{H}_{ex}\left(\mathbf{r},t\right)=\frac{2}{\mu_{0}}\left(\nabla\cdot\frac{A\left(\mathbf{r}\right)}{M_{s}^{2}\left(\mathbf{r}\right)}\nabla\right)\mathbf{M}\left(\mathbf{r},t\right)+\frac{2\mathbf{M}\left(\mathbf{r},t\right)}{\mu_{0}M_{s}\left(\mathbf{r}\right)}\left(\nabla\cdot A\left(\mathbf{r}\right)\nabla\right)\frac{1}{M_{s}\left(\mathbf{r}\right)}\\
-\frac{2\mathbf{m}\left(\mathbf{r},t\right)}{\mu_{0}M_{s}^{2}\left(\mathbf{r}\right)}\cdot\left[\left(\nabla\cdot A\left(\mathbf{r}\right)\nabla\right)\frac{\mathbf{m}\left(\mathbf{r},t\right)}{M_{s}\left(\mathbf{r}\right)}\right]\hat{z}.\label{eq:H_ex}
\end{multline}
The exchange field enters the LLG equation only as
a cross product with the magnetization $\mathbf{M}\left(\mathbf{r},t\right)$.
The second term is parallel to $\mathbf{M}\left(\mathbf{r},t\right)$
and thus will not contribute to Eq.~(\ref{eq:LLG}). The third term
of Eq. (\ref{eq:H_ex}), which is proportional to $\mathbf{m}\left(\mathbf{r},t\right)$
and parallel to $M_{s}\left(\mathbf{r}\right)$, will only produce
terms of second order in $\mathbf{m}\left(\mathbf{r},t\right)$ in
Eq. (\ref{eq:LLG}) and can safely be dropped. Therefore, we can approximate
\begin{equation}
\mathbf{H}_{ex}\left(\mathbf{r},t\right)=\frac{2}{\mu_{0}}\left(\nabla\cdot\frac{A\left(\mathbf{r}\right)}{M_{s}^{2}\left(\mathbf{r}\right)}\nabla\right)\mathbf{M}\left(\mathbf{r},t\right),\label{eq:H_ex_simple}
\end{equation}
which produces a LLG equation from Eq. (\ref{eq:LLG}) that is correct
to first order in $\mathbf{m}\left(\mathbf{r},t\right)$.

We now have the following equation for the effective field:
\begin{equation}
\mathbf{H}_{eff}\left(\mathbf{r},t\right)=H_{0}\hat{z}+\mathbf{h}\left(\mathbf{r},t\right)+\frac{2}{\mu_{0}}\left(\nabla\cdot\frac{A\left(\mathbf{r}\right)}{M_{s}^{2}\left(\mathbf{r}\right)}\nabla\right)\mathbf{M}\left(\mathbf{r},t\right).\label{eq:H_eff}
\end{equation}
This form is a generalization of the boundary condition obtained at the interface between a ferromagnet and vacuum\cite{Rado1959}, in the absence of any surface torque, and later derived for the boundary condition between dissimilar magnetic materials\cite{Hoffmann1970a,Hoffmann1970b}. It is also the form used in Ref.~\onlinecite{Krawczyk2008}.

\subsection{Plane-Wave Solution to LLG Equation for Quasi-Two-Dimensional-Magnonic
Crystal}

When solving for magnons of a specific frequency $\omega$ we write $\mathbf{m}\left(\mathbf{r},t\right) = \mathbf{m}\left(\mathbf{r}\right)\exp{(-i\omega t)}$ and the dipolar field, $\mathbf{h}\left(\mathbf{r},t\right)=-\nabla\Psi\left(\mathbf{r}\right)\exp{(-i\omega t)}$,
with $\Psi\left(\mathbf{r}\right)$ the magnetostatic potential.
With the form of the effective field in Eq. (\ref{eq:H_eff}), the
LLG equation (Eq. (\ref{eq:LLG})) can be written
\begin{eqnarray}
i\Omega m_{x}\left(\mathbf{R}\right)+M_{s}\left(\mathbf{R}\right)\nabla\cdot\left(Q\left(\mathbf{x}\right)\nabla m_{y}\left(\mathbf{R}\right)\right)-m_{y}\left(\mathbf{R}\right)\nabla\cdot\left(Q\left(\mathbf{R}\right)\nabla M_{s}\left(\mathbf{R}\right)\right)\nonumber \\
-m_{y}\left(\mathbf{R}\right)-\frac{M_{s}\left(\mathbf{R}\right)}{H_{0}}\frac{\partial\Psi\left(\mathbf{R}\right)}{\partial y}+i\Omega\alpha\left(\mathbf{R}\right)m_{y}\left(\mathbf{R}\right) & = & 0,\label{eq:LLG_1}\\
i\Omega m_{y}\left(\mathbf{R}\right)-M_{s}\left(\mathbf{R}\right)\nabla\cdot\left(Q\left(\mathbf{R}\right)\nabla m_{x}\left(\mathbf{R}\right)\right)+m_{x}\left(\mathbf{R}\right)\nabla\cdot\left(Q\left(\mathbf{R}\right)\nabla M_{s}\left(\mathbf{R}\right)\right)\nonumber \\
+m_{x}\left(\mathbf{R}\right)+\frac{M_{s}\left(\mathbf{R}\right)}{H_{0}}\frac{\partial\Psi\left(\mathbf{R}\right)}{\partial x}-i\Omega\alpha\left(\mathbf{R}\right)m_{x}\left(\mathbf{R}\right) & = & 0,\label{eq:LLG_2}
\end{eqnarray}
where $\Omega=\omega/(\left|\gamma\right|\mu_{0}H_{0})$ and $Q\left(\mathbf{R}\right)=2A\left(\mathbf{R}\right)/\left(\mu_{0}H_{0}M_{s}^{2}\left(\mathbf{R}\right)\right)$.
Additionally, since there is no $z$ dependence in the above equations,
the three dimensional position vector $\mathbf{r}$ has been replaced
with the two dimensional position vector, $\mathbf{R}=(x,y)$.

This system of equations can be efficiently solved with a plane-wave
method\cite{Vasseur1996,Krawczyk2008,Tiwari2010}.
We take advantage of the crystal's periodicity and use
Bloch's theorem to write the magnetization and magnetostatic potential
as an expansion of plane waves:
\begin{eqnarray}
\mathbf{m}\left(\mathbf{R}\right)=e^{i\mathbf{k}\cdot\mathbf{R}}\sum_{i}\mathbf{m}_{\mathbf{k}}\left(\mathbf{G}_{i}\right)e^{i\mathbf{G}_{i}\cdot\mathbf{R}},\label{eq:Bloch1}\\
\Psi\left(\mathbf{R}\right)=e^{i\mathbf{k}\cdot\mathbf{R}}\sum_{i}\Psi_{\mathbf{k}}\left(\mathbf{G}_{i}\right)e^{i\mathbf{G}_{i}\cdot\mathbf{R}}.\label{eq:Bloch2}
\end{eqnarray}
Here $\mathbf{G}_{i}$ represents a two dimensional reciprocal lattice
vector of the crystal and $\mathbf{k}$ is a wave vector in the first
Brillouin zone. The magnetostatic potential can be rewritten in terms
of the magnetization by using one of Maxwell's equations:
\begin{equation}
\nabla\cdot\left(\mathbf{h}\left(\mathbf{R}\right)+\mathbf{m}\left(\mathbf{R}\right)\right)=0.
\end{equation}
Replacing $\mathbf{h}\left(\mathbf{R}\right)$ with $-\nabla\Psi\left(\mathbf{R}\right)$,
substituting in Eqs. (\ref{eq:Bloch1}) and (\ref{eq:Bloch2}), and solving for the potential
yields
\begin{equation}
\Psi\left(\mathbf{G}\right)=-i\frac{m_{x,\mathbf{k}}\left(\mathbf{G}\right)\left(\mathbf{G}_{x}+k_{x}\right)+m_{y,\mathbf{k}}\left(\mathbf{G}_{y}+k_{y}\right)}{\left(\mathbf{G}+\mathbf{k}\right)^{2}}.
\end{equation}
Next, we need to be able to write the material properties $M_{s}\left(\mathbf{R}\right)$,
$Q\left(\mathbf{R}\right)$, and $\alpha\left(\mathbf{R}\right)$
in reciprocal space. Since these have the same periodicity as the
crystal lattice, this can be done with a Fourier series expansion:
\begin{eqnarray}
M_{s}\left(\mathbf{R}\right) & = & \sum_{i}M_{s}\left(\mathbf{G}_{i}\right)e^{i\mathbf{G}_{i}\cdot\mathbf{R}},\nonumber \\
Q\left(\mathbf{R}\right) & = & \sum_{i}Q\left(\mathbf{G}_{i}\right)e^{i\mathbf{G}_{i}\cdot\mathbf{R},}\label{eq:Fourier}\\
\alpha\left(\mathbf{R}\right) & = & \sum_{i}\alpha\left(\mathbf{G}_{i}\right)e^{i\mathbf{G}_{i}\cdot\mathbf{R}}.\nonumber 
\end{eqnarray}
The Fourier coefficients are obtained by an inverse Fourier transform:
\begin{eqnarray}
M_{s}\left(\mathbf{G}\right) & = & \frac{1}{S}\int_{S}M_{s}\left(\mathbf{R}\right)e^{-i\mathbf{G}\cdot\mathbf{R}}d^{2}\mathbf{R}.
\end{eqnarray}
where S is the area of the two-dimensional unit cell. Performing the
integration for $\mathbf{G}=0$ gives the average
\begin{equation}
M_{s}\left(\mathbf{G}=0\right)=M_{s_{A}}f+M_{s_{B}}\left(1-f\right),
\end{equation}
where $f$ is the fractional space occupied by a cylinder in the unit
cell. For $\mathbf{G}\neq0$, we have 
\begin{equation}
M_{s}\left(\mathbf{G}\neq0\right)=\left(M_{s_{A}}-M_{s_{B}}\right)2f\frac{J_{1}\left(\left|\mathbf{G}\right|R_{cyl}\right)}{\left|\mathbf{G}\right|R_{cyl}}.
\end{equation}
Here $J_{1}$ is a Bessel function of the first kind, and $R_{cyl}$
is the radius of the cylinders. The following infinite system of equations
in reciprocal space is obtained by substituting Eqs. (\ref{eq:Bloch1})-(\ref{eq:Fourier})
in Eqs. (\ref{eq:LLG_1}) and (\ref{eq:LLG_2}):
\begin{multline}
i\Omega\sum_{j}\left(m_{x,\mathbf{k}}\left(\mathbf{G}_{i}\right)\delta_{ij}+\alpha\left(\mathbf{G}_{i}-\mathbf{G}_{j}\right)m_{y,\mathbf{k}}\left(\mathbf{G}_{j}\right)\right)=\\
\sum_{j}\left\{ M_{s}\left(\mathbf{G}_{i}-\mathbf{G}_{j}\right)\frac{\left(G_{x,j}+k_{x}\right)\left(G_{y,j}+k_{y}\right)}{H_{0}\left(\mathbf{G}_{j}+\mathbf{k}\right)^{2}}m_{x,\mathbf{k}}\left(\mathbf{G}_{j}\right)+\left[\delta_{ij}+M_{s}\left(\mathbf{G}_{i}-\mathbf{G}_{j}\right)\frac{\left(G_{y,j}+k_{y}\right)^{2}}{H_{0}\left(\mathbf{G}_{j}+\mathbf{k}\right)^{2}}\right.\right.\\
\left.\left.+\sum_{l}\left(M_{s}\left(\mathbf{G}_{i}-\mathbf{G}_{l}\right)Q\left(\mathbf{G}_{l}-\mathbf{G}_{j}\right)\left(\left(\mathbf{k}+\mathbf{G}_{j}\right)\cdot\left(\mathbf{k}+\mathbf{G}_{l}\right)-\left(\mathbf{G}_{i}-\mathbf{G}_{j}\right)\cdot\left(\mathbf{G}_{i}-\mathbf{G}_{l}\right)\right)\right)\right]m_{y,\mathbf{k}}\left(\mathbf{G}_{j}\right)\right\} 
\end{multline}
\begin{multline}
i\Omega\sum_{j}\left(m_{y,\mathbf{k}}\left(\mathbf{G}_{i}\right)\delta_{ij}-\alpha\left(\mathbf{G}_{i}-\mathbf{G}_{j}\right)m_{x,\mathbf{k}}\left(\mathbf{G}_{j}\right)\right)=\\
-\sum_{j}\left\{ M_{s}\left(\mathbf{G}_{i}-\mathbf{G}_{j}\right)\frac{\left(G_{x,j}+k_{x}\right)\left(G_{y,j}+k_{y}\right)}{H_{0}\left(\mathbf{G}_{j}+\mathbf{k}\right)^{2}}m_{y,\mathbf{k}}\left(\mathbf{G}_{j}\right)+\left[\delta_{ij}+M_{s}\left(\mathbf{G}_{i}-\mathbf{G}_{j}\right)\frac{\left(G_{y,j}+k_{y}\right)^{2}}{H_{0}\left(\mathbf{G}_{j}+\mathbf{k}\right)^{2}}\right.\right.\\
\left.\left.+\sum_{l}\left(M_{s}\left(\mathbf{G}_{i}-\mathbf{G}_{l}\right)Q\left(\mathbf{G}_{l}-\mathbf{G}_{j}\right)\left(\left(\mathbf{k}+\mathbf{G}_{j}\right)\cdot\left(\mathbf{k}+\mathbf{G}_{l}\right)-\left(\mathbf{G}_{i}-\mathbf{G}_{j}\right)\cdot\left(\mathbf{G}_{i}-\mathbf{G}_{l}\right)\right)\right)\right]m_{x,\mathbf{k}}\left(\mathbf{G}_{j}\right)\right\} .
\end{multline}

We  solve this by limiting the number of reciprocal lattice vectors
in the sum and expressing it as a matrix equation:
\begin{equation}
i\Omega\begin{bmatrix}\delta_{ij} & \alpha\left(\mathbf{G}_{i}-\mathbf{G}_{j}\right)\\
\alpha\left(\mathbf{G}_{i}-\mathbf{G}_{j}\right) & \delta_{ij}
\end{bmatrix}\begin{bmatrix}m_{x,\mathbf{k}}\left(\mathbf{G}_{1}\right)\\
\vdots\\
m_{x,\mathbf{k}}\left(\mathbf{G}_{N}\right)\\
m_{y,\mathbf{k}}\left(\mathbf{G}_{1}\right)\\
\vdots\\
m_{y,\mathbf{k}}\left(\mathbf{G}_{N}\right)
\end{bmatrix}=\begin{bmatrix}B_{ij}^{xx} & B_{ij}^{xy}\\
B_{ij}^{yx} & B_{ij}^{yy}
\end{bmatrix}\begin{bmatrix}m_{x,\mathbf{k}}\left(\mathbf{G}_{1}\right)\\
\vdots\\
m_{x,\mathbf{k}}\left(\mathbf{G}_{N}\right)\\
m_{y,\mathbf{k}}\left(\mathbf{G}_{1}\right)\\
\vdots\\
m_{y,\mathbf{k}}\left(\mathbf{G}_{N}\right)
\end{bmatrix}\label{eq:Matrix Eqn}
\end{equation}
\begin{eqnarray}
B_{ij}^{xx} & = & -B_{ij}^{yy}=M_{s}\left(\mathbf{G}_{i}-\mathbf{G}_{j}\right)\frac{\left(G_{x,j}+k_{x}\right)\left(G_{y,j}+k_{y}\right)}{H_{0}\left(\mathbf{G}_{j}+\mathbf{k}\right)^{2}}\\
B_{ij}^{xy} & = & \delta_{ij}+M_{s}\left(\mathbf{G}_{i}-\mathbf{G}_{j}\right)\frac{\left(G_{y,j}+k_{y}\right)^{2}}{H_{0}\left(\mathbf{G}_{j}+\mathbf{k}\right)^{2}}\nonumber \\
 &  & +\sum_{l}M_{s}\left(\mathbf{G}_{i}-\mathbf{G}_{l}\right)Q\left(\mathbf{G}_{l}-\mathbf{G}_{j}\right)\left[\left(\mathbf{k}+\mathbf{G}_{j}\right)\cdot\left(\mathbf{k}+\mathbf{G}_{l}\right)-\left(\mathbf{G}_{i}-\mathbf{G}_{j}\right)\cdot\left(\mathbf{G}_{i}-\mathbf{G}_{l}\right)\right]\\
B_{ij}^{yx} & = & =\delta_{ij}+M_{s}\left(\mathbf{G}_{i}-\mathbf{G}_{j}\right)\frac{\left(G_{x,j}+k_{x}\right)^{2}}{H_{0}\left(\mathbf{G}_{j}+\mathbf{k}\right)^{2}}\nonumber \\
 &  & +\sum_{l}M_{s}\left(\mathbf{G}_{i}-\mathbf{G}_{l}\right)Q\left(\mathbf{G}_{l}-\mathbf{G}_{j}\right)\left[\left(\mathbf{k}+\mathbf{G}_{j}\right)\cdot\left(\mathbf{k}+\mathbf{G}_{l}\right)-\left(\mathbf{G}_{i}-\mathbf{G}_{j}\right)\cdot\left(\mathbf{G}_{i}-\mathbf{G}_{l}\right)\right].
\end{eqnarray}
The LLG equation is now reduced to finding the eigenvalues and eigenvectors
for the above equation.

\section{Results\label{results}}


From Eq. (\ref{eq:Matrix Eqn}) we  calculate the complex eigenvalues
$\Omega$ corresponding to the frequencies of magnons in the two-dimensional magnetic superlattices of Fe, Co, Ni, and YIG. The real part of $\Omega_n(\mathbf{k})$ is the magnon frequency of branch $n$
for the wave vector $\mathbf{k}$ and the imaginary part is
the inverse spin wave lifetime.  To focus on the dependence of these properties on magnetic material combinations we consider superlattices with a lattice constant $a=10\mbox{nm}$, an
external field $\mu_{0}H_{0}=0.1\mbox{T}$, and a filling fraction
$f=0.5$. The material properties, $M_{s}$, $A$, and $\alpha$,
are listed in Table \ref{Table: Material Properties}.

\begin{figure}[H]
\begin{center}

\includegraphics{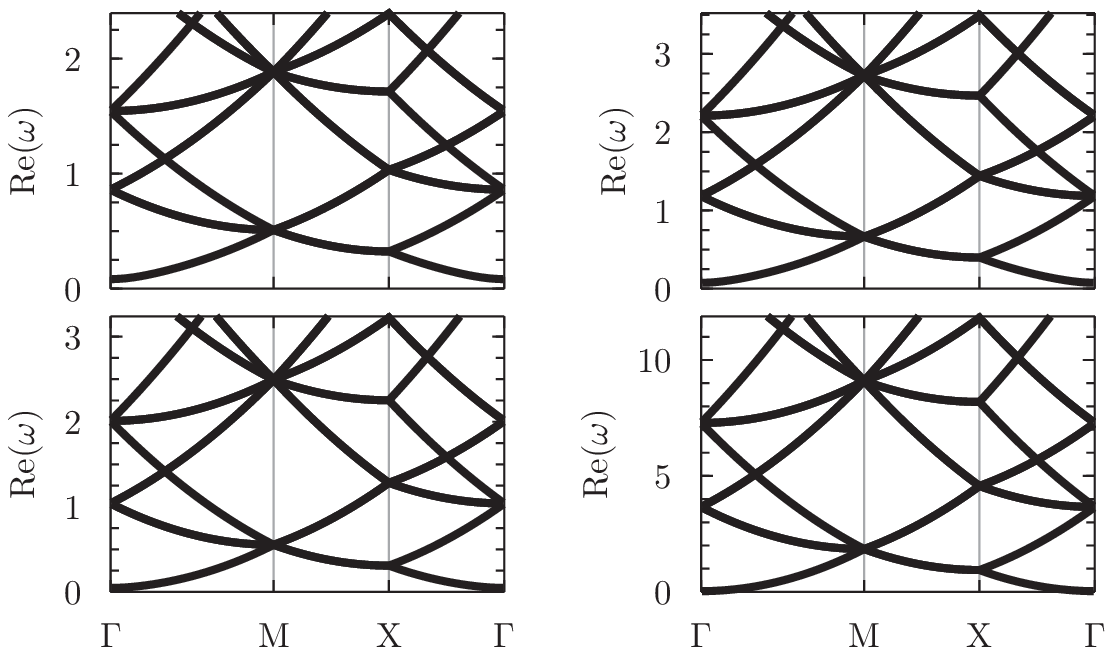}

\end{center}\caption{Empty square lattice band structure obtained from the LLG equation
for a homogeneous crystal of Fe (upper left), Co (upper right), Ni
(lower left), and YIG (lower right) with lattice constant $a=10\mbox{nm}$. Frequencies are in units of THz.}
 \label{Fig: Hom Square Band Structure}
\end{figure}

\begin{figure}[H]
\begin{center}

\includegraphics{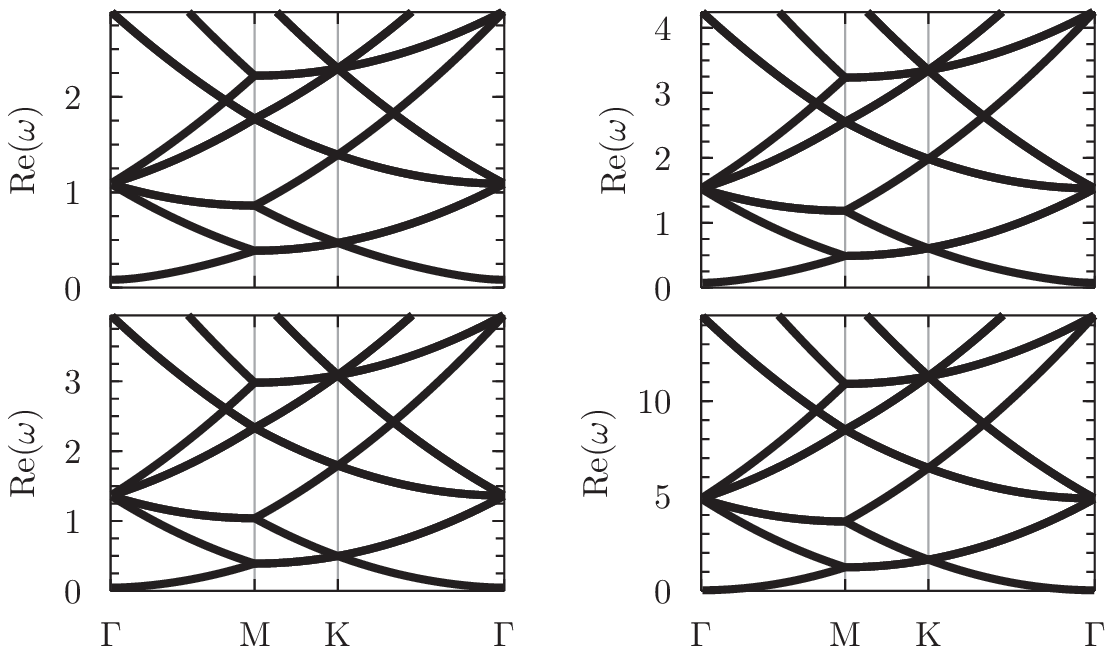}

\end{center}

\caption{Empty hexagonal lattice band structure obtained from the LLG equation
for a homogeneous crystal of Fe (upper left), Co (upper right), Ni
(lower left), and YIG (lower right) with lattice constant $a=10\mbox{nm}$. Frequencies are in units of THz.}
\label{Fig: Hom Hexagonal Band Structure}
\end{figure}

Figures \ref{Fig: Hom Square Band Structure} and \ref{Fig: Hom Hexagonal Band Structure},
 show the empty-lattice band structures obtained from the LLG equation for homogeneous
crystals of Fe, Co, Ni, and YIG. As the empty-lattice features are governed by the lattice symmetry and the material's spin wave velocity, these plots depend on material only in setting the frequency scale of the features.

\begin{figure}[h]
\begin{center}

\includegraphics{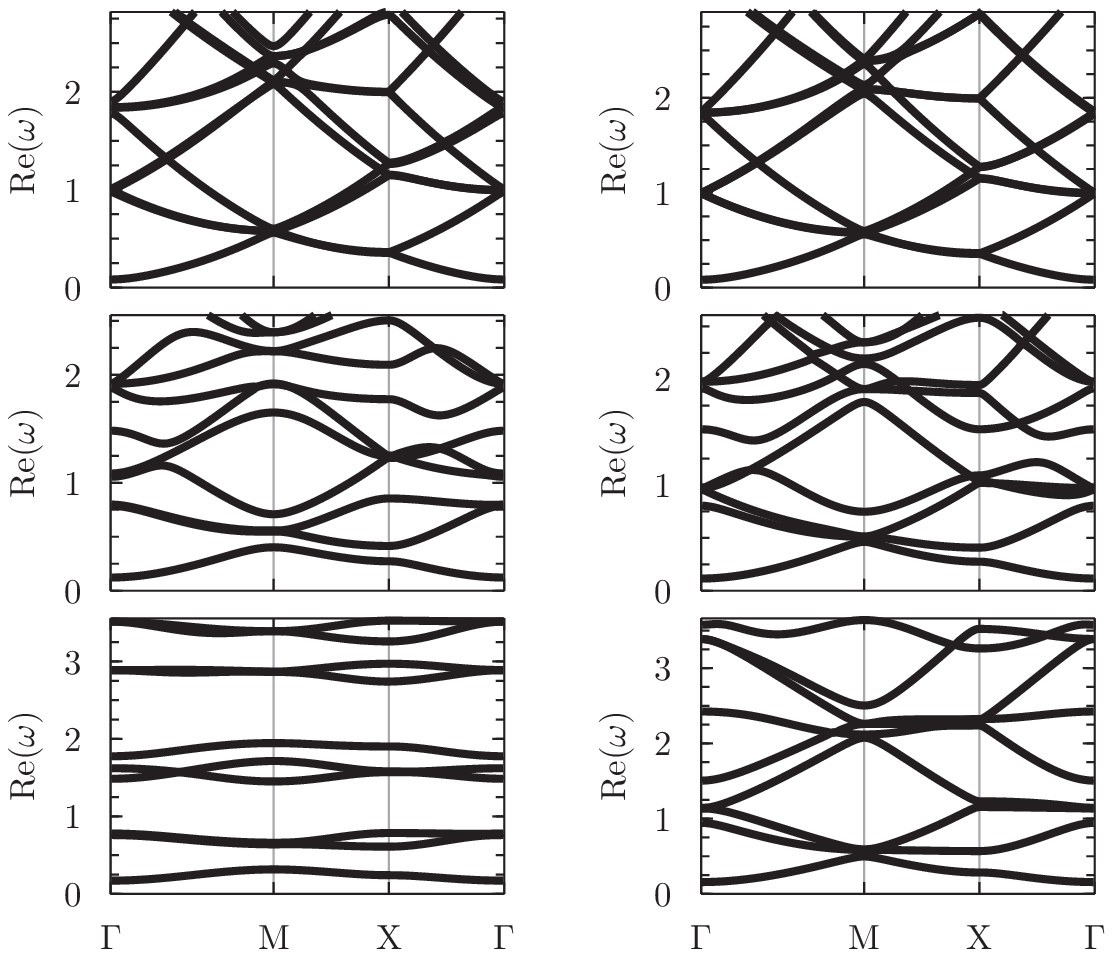}

\end{center}

\caption{Magnonic band structures for a square lattice magnonic crystal with lattice
spacing $a=10\mbox{nm}$ and filling fraction $f=0.5$. On the left
is Fe cylinders embedded in Co (top), Ni (middle), and YIG (bottom).
The right is for an Fe host with Co (top), Ni (middle), and YIG (bottom)
cylinders. Frequencies are in units of THz.}

\label{Fig:Various Square Band Structures}
\end{figure}

\begin{figure}[H]
\begin{center}

\includegraphics{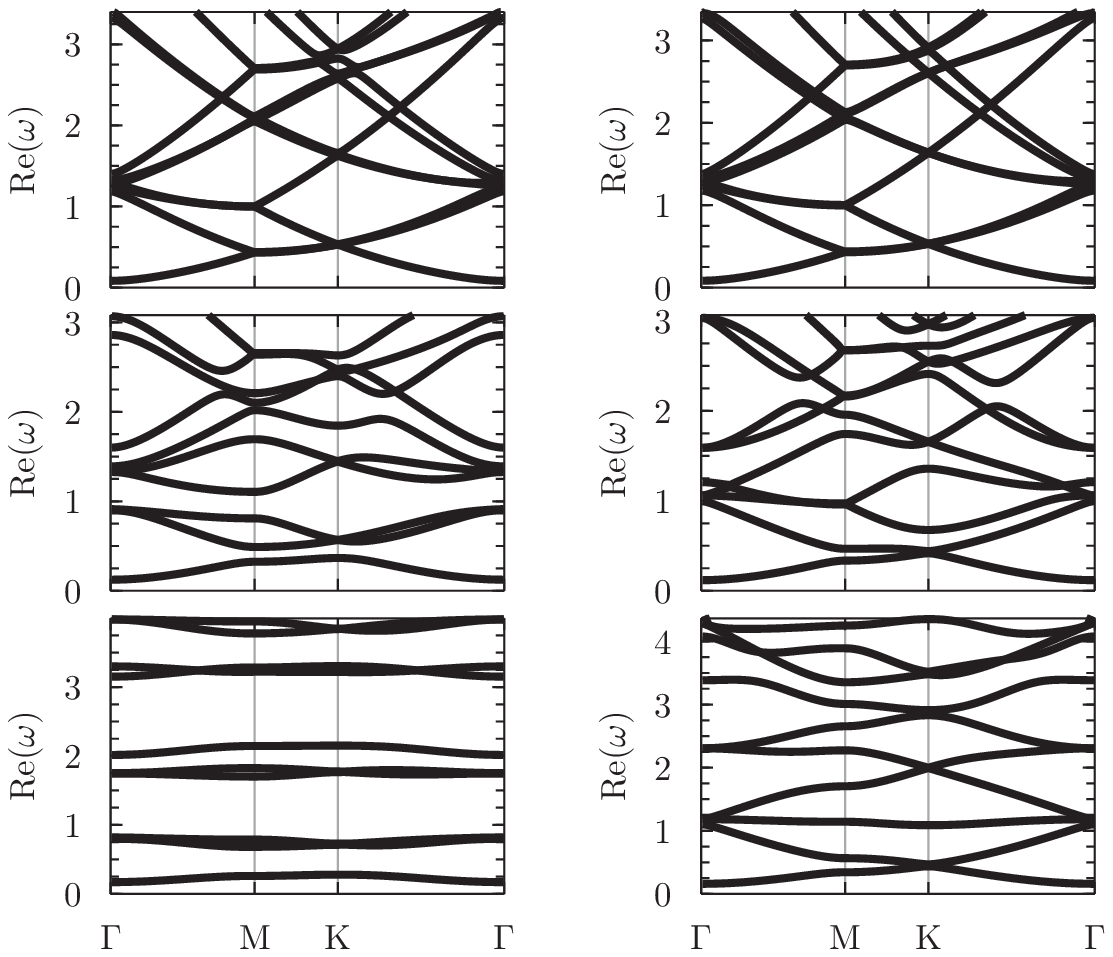}

\end{center}

\caption{Magnonic band structures for a hexagonal lattice magnonic crystal with lattice
spacing $a=10\mbox{nm}$ and filling fraction $f=0.5$. On the left
is Fe cylinders embedded in Co (top), Ni (middle), and YIG (bottom).
The right is for an Fe host with Co (top), Ni (middle), and YIG (bottom)
cylinders. Frequencies are in units of THz.}
\label{Fig: Various Hexagonal Band Structures}
\end{figure}

Figs. \ref{Fig:Various Square Band Structures} and
\ref{Fig: Various Hexagonal Band Structures} show the results for
when Fe is combined with Co, Ni, or YIG. The change
in band structure from the homogeneous case is more substantial when
there is a greater difference in the spontaneous magnetization between
the two materials. For example, the magnetics properties of Fe and
Co are fairly similar, and so for a crystal composed of these materials,
the band structure differs little from the homogeneous case, with
only some small splittings of the spin wave dispersion curves occurring. However,
when for a crystal of Fe and YIG, whose magnetizations differ
by more than a factor of ten, the magnonic modes are almost completely
different from the homogeneous crystal. Furthermore, the
opening of band gaps in these structures is more easily attainable
when the magnetization is larger in the cylinders than it is in the
host. With Fe cylinders embedded in YIG in a square lattice, there
are four gaps occurring within the lowest nine spin wave modes, while
YIG cylinders in Fe shows only one small gap between the first and
second spin wave modes. 

\begin{figure}[H]
\begin{center}
\includegraphics[scale=0.4]{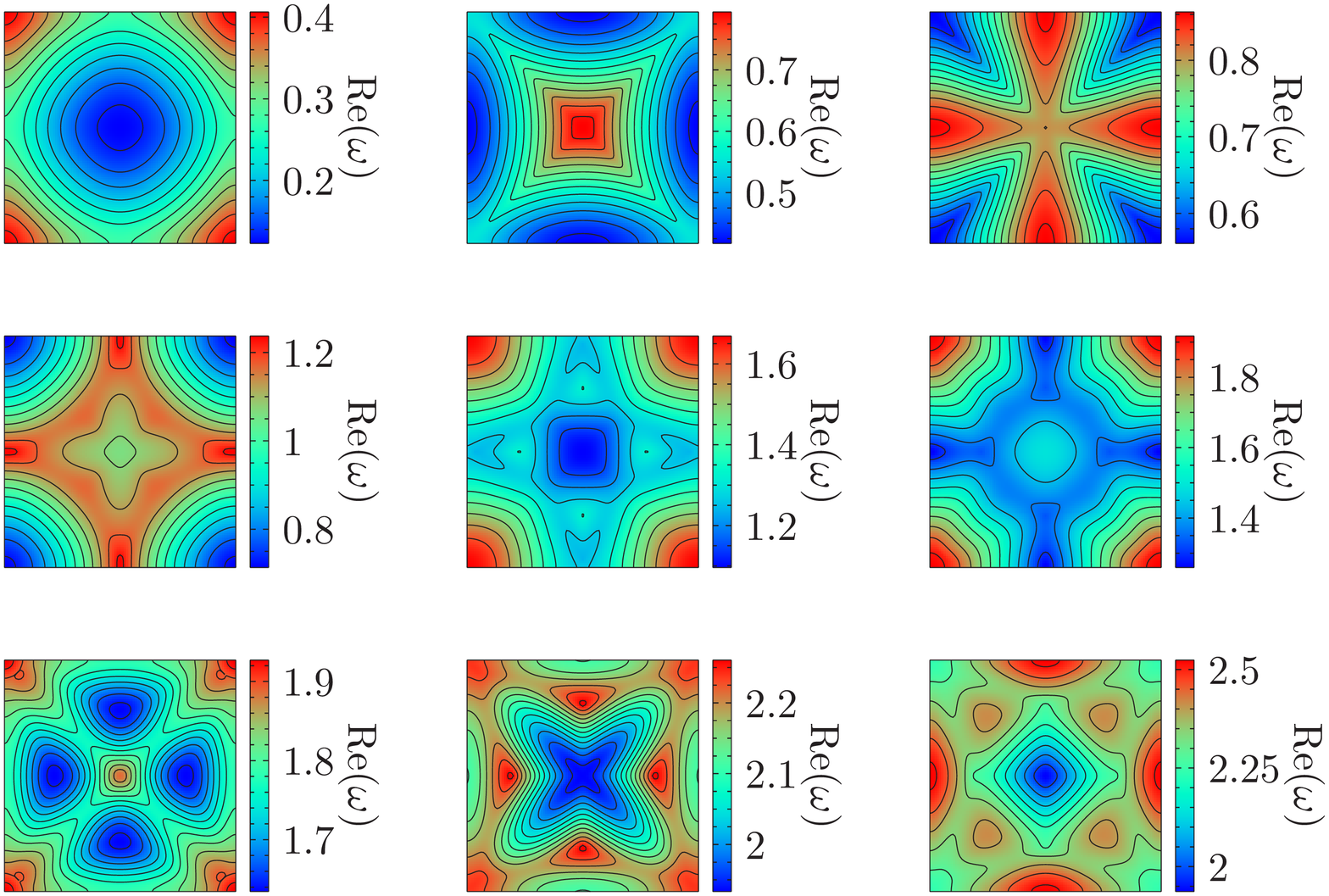}
\end{center}\caption{(Color online) The lowest nine spin wave frequencies (in THz) for a square lattice magnonic crystal
composed of Fe cylinders in Ni with a filling fraction of $f=0.5$,
and a lattice constant of $a=10\mbox{nm}$. These results were obtained
using the exchange field in Eq. (\ref{eq:H_ex_simple}).}
\label{Fig: Contours Real}
\end{figure}

\begin{figure}[h]
\begin{center}

\includegraphics[scale=0.4]{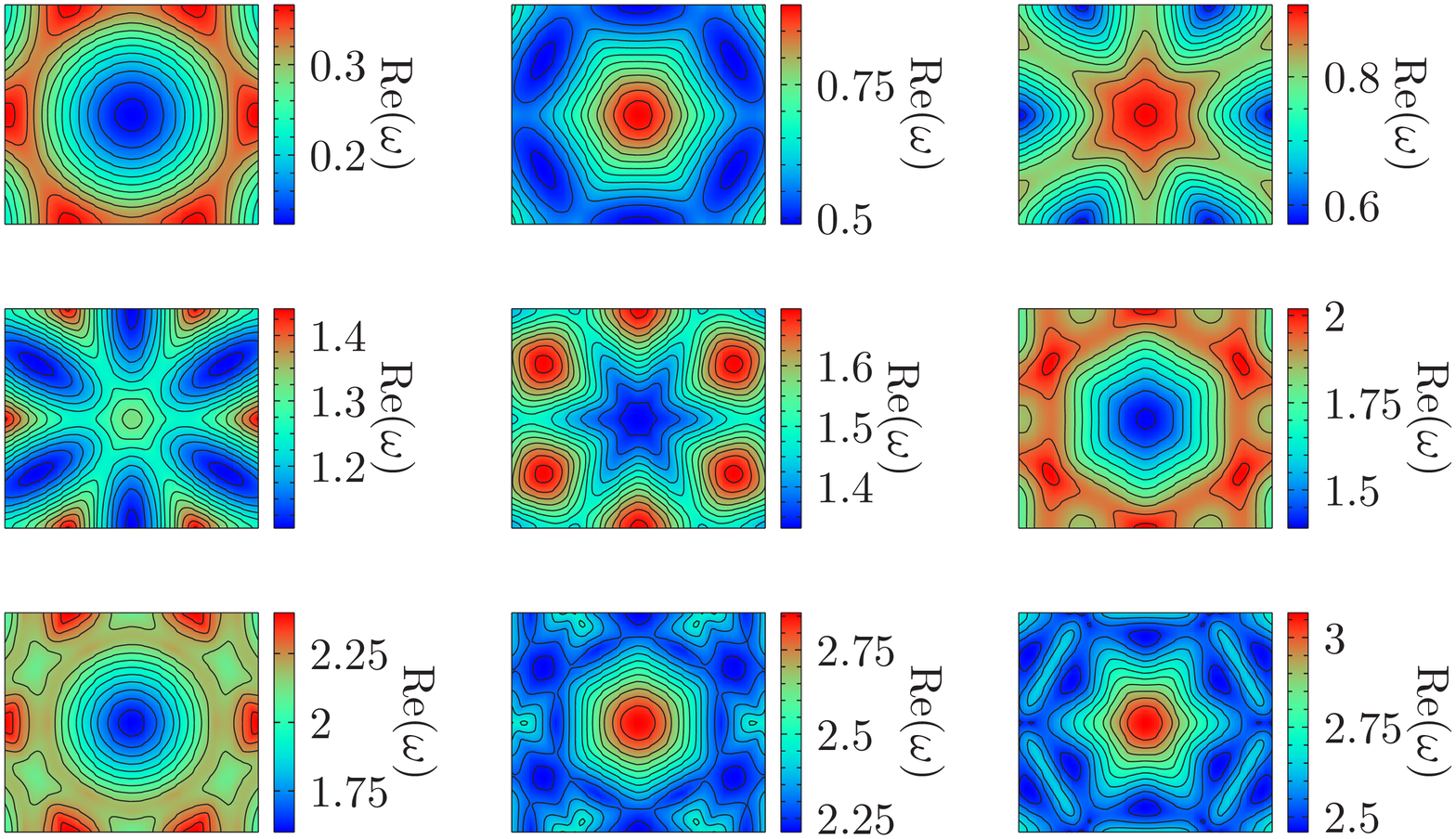}

\end{center}

\caption{(Color online) The lowest nine spin wave frequencies (in THz) for a hexagonal lattice magnonic crystal
composed of Fe cylinders in Ni with a filling fraction of $f=0.5$,
and a lattice constant of $a=10\mbox{nm}$. These results were obtained
using the exchange field in Eq. (\ref{eq:H_ex_simple}).}
\label{Fig: Hex Contours Real}
\end{figure}

\begin{figure}[h]
\begin{center}

\includegraphics[scale=0.4]{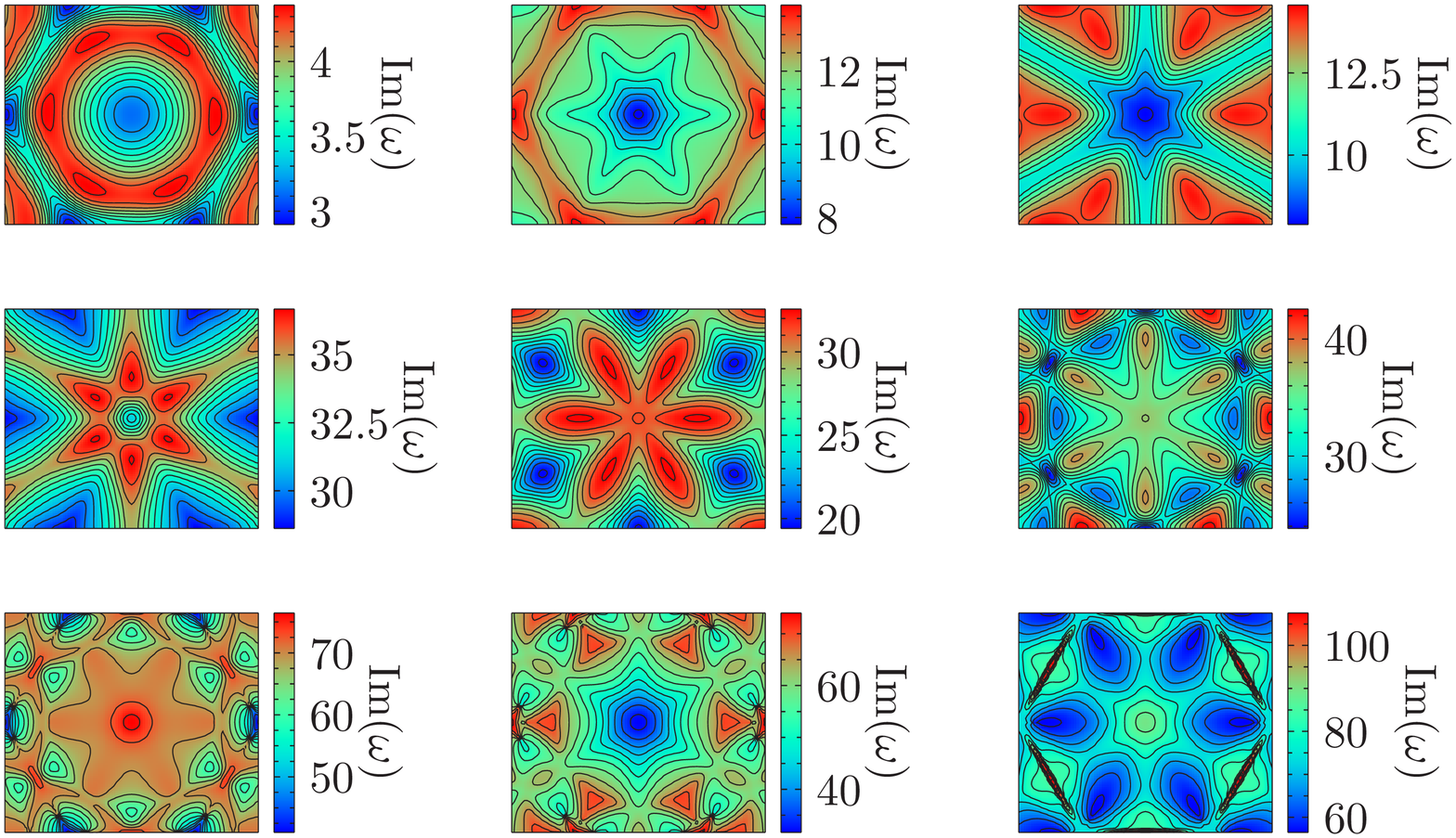}

\end{center}

\caption{(Color online) The  spin wave relaxation rates (in units of GHz) corresponding to the lowest nine spin wave modes from Fig.~\ref{Fig: Hex Contours Real}}
\label{Fig: Hex Contours Imag}
\end{figure}

 Figs. \ref{Fig: Contours Real}-\ref{Fig: Hex Contours Imag} show the detailed dispersion curves and spin wave relaxation rates for a hexagonal superlattice of Fe cylinders in Ni. Plotted are the  lowest nine spin wave modes (${\rm Re}\left(\Omega\right)$)
in the entire first Brilllouin zone as well as the corresponding inverse
spin wave lifetimes (${\rm Im}\left(\Omega\right)$). Quality factors for these modes, corresponding to the ratio of the relaxation rate to the mode frequency, can exceed 100 for such spin waves, especially for the lowest-frequency modes.

\section{Comparison with alternate effective field\label{comparison}}

\begin{figure}[h]
\begin{center}

\includegraphics{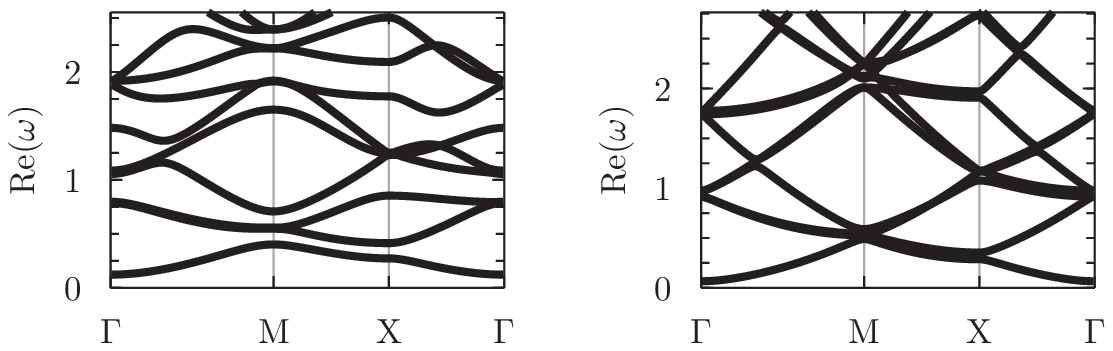}

\end{center}\caption{Magnonic band structure of Fe cylinders embedded in Ni calculated from the
LLG equation using exchange field in Eq. (\ref{eq:H_ex_simple}) (left)
and Eq. (\ref{eq:H_ex_Alt}) (right). Frequencies are in units of THz.}
\label{Fig: Band Structures (Right/Wrong)}
\end{figure}

Some recent calculations of  magnonic crystals dispersion curves used a different exchange field\cite{Vasseur1996,Tiwari2010}
than the one derived in Sec.~\ref{formalism}. The alternate
form,
\begin{equation}
\mathbf{H}_{ex}\left(\mathbf{r},t\right)=\frac{2}{\mu_{0}M_{s}\left(\mathbf{r}\right)}\left(\nabla\cdot\frac{A\left(\mathbf{r}\right)}{M_{s}\left(\mathbf{r}\right)}\nabla\right)\mathbf{M}\left(\mathbf{r},t\right).\label{eq:H_ex_Alt}
\end{equation}
differs by the positioning of one factor of $M_{s}\left(\mathbf{r}\right)^{-1}$ outside the gradient operators. 
A comparison of the band structures obtained for the two different
exchange fields is shown in Fig. \ref{Fig: Band Structures (Right/Wrong)}.
An examination of the band structure for a homogeneous material
composed of Fe or Ni (Fig. \ref{Fig: Hom Square Band Structure}) indicates that the results for the derived exchanged field (Eq. (\ref{eq:H_ex_simple}))
produce a band structure that is appreciably different from the homogeneous
case, whereas the band structures produced by the alternate
exchange field ( Eq. (\ref{eq:H_ex_Alt})) are very similar to the
homogeneous crystal.

\begin{figure}[H]
\begin{center}

\includegraphics[scale=0.4]{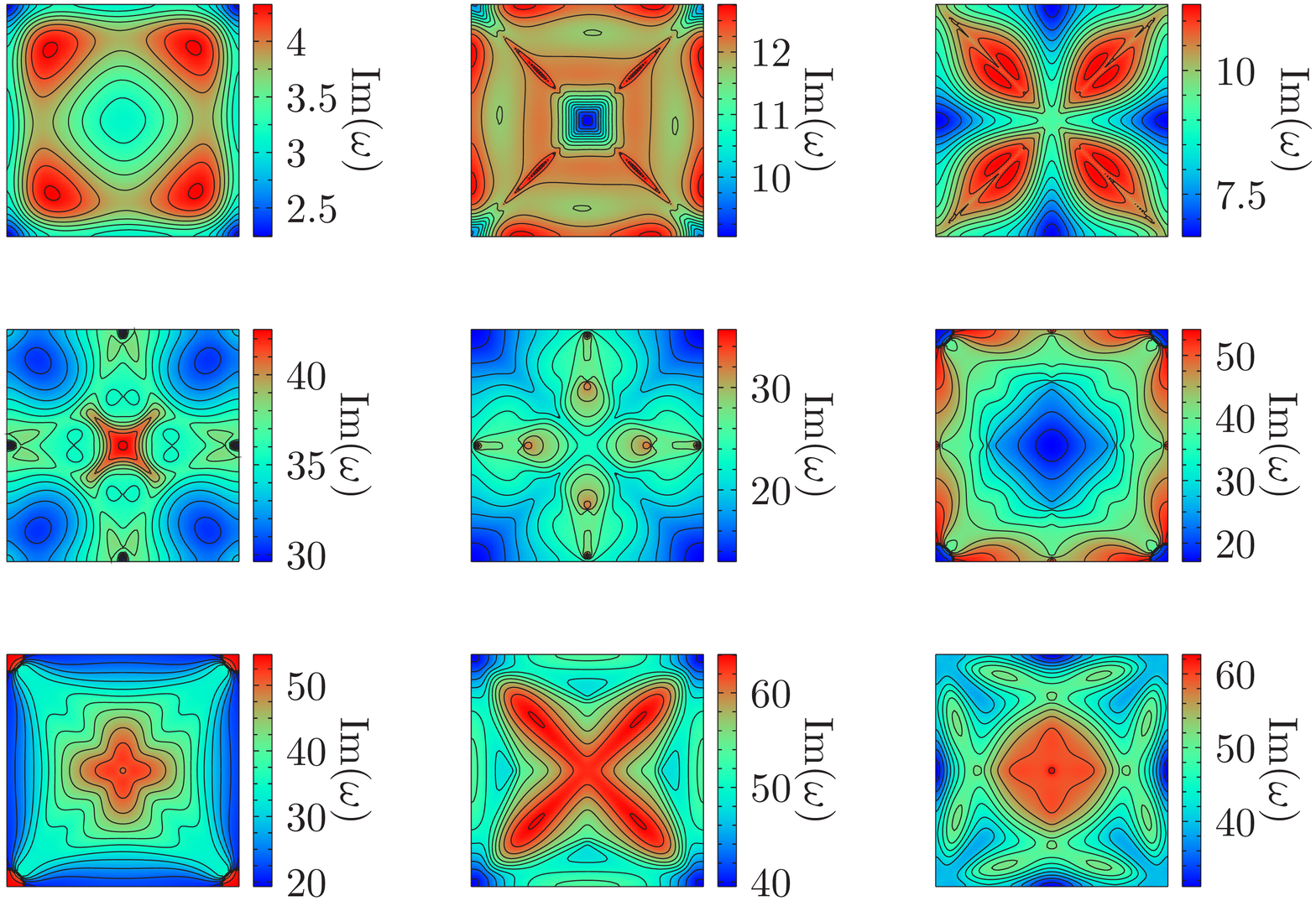}

\end{center}\caption{(Color online) The  spin wave relaxation rate (in units of GHz) corresponding to the lowest nine spin wave modes from Fig.~\ref{Fig: Contours Real}. }
\label{Fig: Contours Imag}
\end{figure}

\begin{figure}[H]
\begin{center}

\includegraphics[scale=0.5]{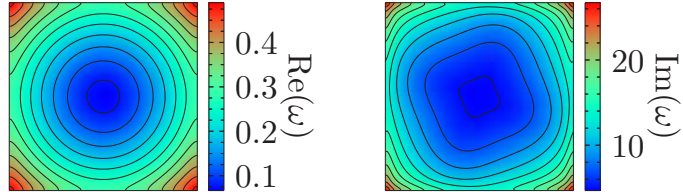}

\end{center}

\caption{(Color online) The lowest spin wave mode (in THz) and the corresponding relaxation rate (in units of GHz)
for a square lattice magnonic crystal composed of Fe cylinders in
Ni with a filling fraction of $f=0.5$, and a lattice constant of
$a=10\mbox{nm}$. These results were obtained using the exchange field
in Eq. (\ref{eq:H_ex_Alt}).}
\label{Fig: Contours Wrong}
\end{figure}

In Fig. \ref{Fig: Contours Wrong} we show the lowest spin wave mode
and corresponding relaxation rate obtained for Fe cylinders
in Ni when using Eq. (\ref{eq:H_ex_Alt}) as the exchange field. When
looking at these contours, we would expect them to have the same symmetries
as the real space lattice. For a square lattice, that would be symmetry
under rotations of $90^{\circ}$ and symmetry under reflections about
either axis. All symmetries are  present for the spin wave modes
of Fig. \ref{Fig: Contours Wrong}, however, the spin wave lifetimes
are lacking the reflection symmetries. Comparison with Figs. \ref{Fig: Contours Real}
and \ref{Fig: Contours Imag} shows that the use of Eq. \ref{eq:H_ex_simple}
for the exchange field keeps the symmetries of the lattice preserved
in the contours. A very slight asymmetry  in Figs. \ref{Fig: Contours Imag} and \ref{Fig: Hex Contours Imag}
is caused by terminating the infinite summation of
the reciprocal lattice vectors in the LLG equation (Eq. (\ref{eq:Matrix Eqn})), and disappears as the number of reciprocal lattice vectors is increased; the asymmetry in Fig.~\ref{Fig: Contours Wrong} does not.

Thus the consequences of using the exchange field of  Eq. (\ref{eq:H_ex_Alt}) instead of the correct form, 
Eq. (\ref{eq:H_ex_simple}), include a dramatic underestimate of the splittings in the magnonic crystal dispersion relations, as well as flawed rotational symmetry of the spin wave relaxation rate.

\section{Conclusion}

Spin wave dispersion curves and relaxation rates have been calculated for hexagonal and square two-dimensional superlattices of magnetic cylinders embedded in another magnetic material. The correct form of the exchange field at the boundary between these two magnetic materials has been found, and the difference from another form used in the literature has been shown to be significant. Full-zone magnonic gaps are obtained for superlattice materials that differ substantially in their saturation magnetization, such as Fe and YIG. Quality factors for spin waves can exceed 100, especially for the lowest-frequency spin mode. These results should assist in the design of magnonic crystals that can focus or redirect spin waves due to their effective band structure.

\section*{Acknowledgments} We acknowledge helpful conversations with A. D. Kent and F. Macia and support from an ARO MURI.
\vfill\eject

\bibliographystyle{apsrev4-1}
\bibliography{central-bibliography}
\vfill\eject
\begin{table}[h]
\begin{tabular}{|c|c|c|c|}
\hline 
 & $M_{s}\mbox{(A/m)}$ & $A\mbox{(pJ/m)}$ & $\alpha$\tabularnewline
\hline 
\hline 
Fe & $1.711\cdot10^{6}$ & $8.3$ & $0.0019$\tabularnewline
\hline 
Co & $1.401\cdot10^{6}$ & $10.3$ & $0.011$\tabularnewline
\hline 
Ni & $0.485\cdot10^{6}$ & $3.4$ & $0.064$\tabularnewline
\hline 
YIG & $0.14\cdot10^{6}$ & $4.15$ & $0.0014$\tabularnewline
\hline 
\end{tabular}\caption{Properties of the different materials considered for the magnonic
crystals\cite{Skomski2006,Oogane2006,Slonczewski1974,Zhang1996}.}
\label{Table: Material Properties}
\end{table}

\vfill\eject




\end{document}